\documentstyle[12pt]{article}
\newcommand {\beq}{\begin{equation}}
\newcommand {\eeq}{\end{equation}}
\newcommand {\beqa}{\begin{eqnarray}}
\newcommand {\eeqa}{\end{eqnarray}}
\newcommand {\beqan}{\begin{eqnarray*}}
\newcommand {\eeqan}{\end{eqnarray*}}
\newcommand {\n}{\nonumber \\}

\newcommand {\Romannumeral}[1]{\uppercase\expandafter{\romannumeral#1}}

\renewcommand{\theequation}{\thesection.\arabic{equation}}

\begin{document}
\setlength{\oddsidemargin}{0cm}
\setlength{\baselineskip}{7mm}

\begin{titlepage}
 \renewcommand{\thefootnote}{\fnsymbol{footnote}}
$\mbox{ }$
\begin{flushright}
\begin{tabular}{l}
KEK-TH-503 \\
TIT/HEP-357\\
hep-th/9612115\\
December 1996
\end{tabular}
\end{flushright}

~~\\
~~\\
~~\\
\vspace*{0cm}
    \begin{Large}
       \vspace{2cm}
       \begin{center}
         {\Large  A Large-N Reduced Model as Superstring}      \\
       \end{center}
    \end{Large}

  \vspace{1cm}

\begin{center}
           Nobuyuki I{\sc shibashi}$^{1)}$\footnote
           {
e-mail address : ishibash@theory.kek.jp},
           Hikaru K{\sc awai}$^{1)}$\footnote
           {
e-mail address : kawaih@theory.kek.jp},
           Yoshihisa K{\sc itazawa}$^{2)}$\footnote
           {
e-mail address : kitazawa@th.phys.titech.ac.jp}{\sc and}
           Asato T{\sc suchiya}$^{1)}$\footnote
           {e-mail address : tsuchiya@theory.kek.jp,{}~JSPS Research Fellow}\\
      \vspace{1cm}
        $^{1)}$ {\it National Laboratory for High Energy Physics (KEK),}\\
               {\it Tsukuba, Ibaraki 305, Japan} \\
        $^{2)}$ {\it Department of Physics, 
Tokyo Institute of Technology,} \\
                 {\it Oh-okayama, Meguro-ku, Tokyo 152, Japan}\\
\end{center}

\vfill

\begin{abstract}
\noindent 
A matrix model which has the manifest ten-dimensional $N=2$ super Poincare
invariance is proposed. Interactions between BPS-saturated states are analyzed to
show that massless spectrum is the same as that of type IIB string theory. 
It is conjectured
that the large-N reduced model of ten-dimensional super Yang-Mills theory
can be regarded as a constructive definition of this model and therefore is
equivalent to superstring theory.
\end{abstract}
\vfill
\end{titlepage}
\vfil\eject


\section{Introduction}
\setcounter{equation}{0}
\renewcommand{\thefootnote}{\fnsymbol{footnote}}
\hspace*{\parindent}
A number of recent developments on the nonperturbative aspects of string
theory have revealed the fact that various superstring theories 
can be
equivalent even though their original definitions 
look totally different \cite{MS,Witten,Polchinski}. 
It seems that
the universality class of superstring theory is unique so that all
the consistent quantum theories of gravity are equivalent. Contrary to these
remarkable developments, we still have not reached the final stage in 
understanding the string dynamics, and  we definitely need a non-perturbative or
constructive definition of string theory.

In this paper, as a candidate for such constructive definition, 
we would like to propose the large-N reduced model of 
ten-dimensional super Yang-Mills theory \cite{RM}. 
If we assume the existence of its continuum limit we can present
several evidences which indicate that it is equivalent to 
superstring theory and thus gives its constructive definition.
Our work is motivated by the recent work of Banks, Fischler, Shenker
and Susskind \cite{BFSS} in which they propose a matrix model quantum mechanics. 
While their model is relevant 
to a possible nonperturbative definition of M-theory \cite{Witten}, 
our model is directly related to type IIB superstring. 

In section 2 we propose a matrix model which looks like the Green-Schwarz action
of type IIB string \cite{GS} in the Schild gauge \cite{Schild}. 
This matrix model has the manifest
Lorentz invariance and $N=2$ space-time supersymmetry. 
Therefore if the theory is totally
well-defined it will be a constructive definition of string theory. However, this
theory has an infrared divergence and we have to remove it by some 
renormalization, which we discuss in section 4.

In section 3 we consider the one-loop effective action of the model and examine 
the interactions 
between the BPS-saturated objects, and show that the massless spectrum of the 
theory is indeed
consistent with that of type IIB theory. This is another evidence that the matrix
model considered here is equivalent to string theory.

In section 4 we discuss the renormalization of the matrix model and show that
it can be regarded as the continuum limit of the 
large-N reduced model of ten-dimensional
super Yang-Mills theory. Here the old idea of the large-N reduction of
the degrees of freedom plays a rather important role.
In contrast to the non-supersymmetric case,
because of the supersymmetry, the $U(1)^d$ symmetry is not spontaneously 
broken but preserved marginally even in the weak coupling limit. 
Therefore in this limit
the eigenvalues of the gauge fields become free, and they play the role of
the space-time coordinates. It turns out that the space-time is dynamically
generated as a collective coordinate of this model.

Appendix is devoted to show the technical details of the one-loop calculations.

\vspace{1cm}

\section{Relations between Green-Schwarz superstring and a matrix model}
\setcounter{equation}{0}
\hspace*{\parindent}
In this section, we examine the path integral of type IIB string theory
in the Schild gauge and propose a matrix model which can be regarded as its
rigorous definition.

We start with the covariant Green-Schwarz action of type IIB
superstring expressed in the Nambu-Goto form \cite{GS}:
\beqa
S_{GS}=-T\int d^2\sigma &(&
          \sqrt{-\frac{1}{2}\Sigma^2}+i\epsilon^{ab}\partial_a X^{\mu}
          (\bar{\theta}^1\Gamma_{\mu}\partial_b\theta^1
           +\bar{\theta}^2\Gamma_{\mu}\partial_b\theta^2) \n
          & & +\epsilon^{ab}\bar{\theta}^1\Gamma^{\mu}\partial_a\theta^1
              \bar{\theta}^2\Gamma_{\mu}\partial_b\theta^2 ).
\label{SGS}
\eeqa    
Here $\theta^1$ and $\theta^2$ are Majorana-Weyl spinors in ten dimensions
having the same chiralities. $\Sigma^{\mu\nu}$ and $\Pi^{\mu}_a$ are
defined by
\beqa
\Sigma^{\mu\nu}&=&\epsilon^{ab}\Pi^{\mu}_a\Pi^{\nu}_b,  \n
\Pi^{\mu}_a&=&\partial_a X^{\mu}-i\bar{\theta}^1\Gamma^{\mu}\partial_a\theta^1
                             +i\bar{\theta}^2\Gamma^{\mu}\partial_a\theta^2.
\eeqa
Note that we have applied an analytic continuaton in the path integral,
$\theta^2 \rightarrow i \theta^2$, so that the signs of the 
terms including $\theta^2$ are opposite to those of the ordinary ones. 
(See the footnote below.) 
This system possesses the $N=2$ space-time supersymmetry
\beqa
\delta_{SUSY}\theta^1&=& \epsilon^1 ,\n
\delta_{SUSY}\theta^2&=& \epsilon^2 ,\n
\delta_{SUSY} X^{\mu}&=&i\bar{\epsilon}^1\Gamma^{\mu}\theta^1
                      -i\bar{\epsilon}^2\Gamma^{\mu}\theta^2 ,
\label{n2stsusy}
\eeqa
and the $\kappa-$symmetry
\beqa
\delta_{\kappa}\theta^1&=& \alpha^1  ,\n
\delta_{\kappa}\theta^2&=& \alpha^2  ,\n
\delta_{\kappa}X^{\mu}&=&i\bar{\theta}^1\Gamma^{\mu}\alpha^1
                      -i\bar{\theta}^2\Gamma^{\mu}\alpha^2 ,
\eeqa
where
\beqa
\alpha^1&=&(1+\tilde{\Gamma})\kappa^1 ,\n
\alpha^2&=&(1-\tilde{\Gamma})\kappa^2 ,\n
\tilde{\Gamma}&=&\frac{1}{2\sqrt{-\frac{1}{2}\Sigma^2}}
                   \Sigma_{\mu\nu}\Gamma^{\mu\nu}.
\eeqa
Since $\kappa^1$ and $\kappa^2$ are local fermionic parameters and 
$\tilde{\Gamma}^2=1$,
half of the degrees of freedom of $\theta^1$ and $\theta^2$ are redundant, 
and the $\kappa-$symmetry can be gauge-fixed by imposing 
the condition $\theta^1=\theta^2=\psi$. This condition is compatible with 
Lorentz symmetry 
because $\theta^1$ and $\theta^2$ have the same chirality in type IIB string 
theory. 
This gauge-fixing leads to the action\footnote{
Here we comment on the analytic continuation. For an example, we consider
the quantum fluctuations of $\psi$ around the static solution $X^0=\tau, 
X^1=\sigma$ and $X^i=0 \: (i\geq 2)$. Then the action (\ref{SGStilde}) results in
$2iT\int d^2\sigma (\psi^T_R (\partial_\tau+\partial_\sigma)\psi_R-
                   \psi^T_L (\partial_\tau-\partial_\sigma)\psi_L)$.
By redefining $\psi_L$ as $\psi_L\rightarrow -i\psi_L$, 
we obtain the ordinary fermion action
of type IIB superstring. We need this prescription in quantizing the theory
in the hamiltonian formalism. This comes from the analytic continuation,
$\theta^2 \rightarrow i\theta^2$, applied in the path integral quantization of
(\ref{SGS}) 
because the gauge fixing 
leads to $\psi_R=\theta^1_R$ and $\psi_L=\theta^2_L$ in this case.
Therefore what we have done is to analytically continue the degrees of freedom
which is gauged away.
As we see below in the
(\ref{redefN=2}), we also need this kind of prescription naturally in defining 
the supercharges to obtain the correct algebra of $N=2$ supersymmetry.}
\beq
\tilde{S}_{GS}=-T\int d^2\sigma (
          \sqrt{-\frac{1}{2}\sigma^2}+2i\epsilon^{ab}\partial_a X^{\mu}
          \bar{\psi}\Gamma_{\mu}\partial_b\psi) ,
\label{SGStilde}
\eeq
where
\beq
\sigma^{\mu\nu}=\epsilon^{ab}\partial_a X^{\mu} \partial_b X^{\nu}.
\eeq
The action $\tilde{S}_{GS}$ is still invariant under the $N=2$ supersymmetry, 
provided one modifies the transformation law (\ref{n2stsusy}) by mixing it with 
the $\kappa$-symmetry transformation so that the gauge fixing condition 
 $\theta^1=\theta^2$ is preserved. 
The new transformation law becomes,
\beqa
\delta\theta^1&=&\delta_{SUSY}\theta^1+\delta_{\kappa}\theta^1 ,\n
\delta\theta^2&=&\delta_{SUSY}\theta^2+\delta_{\kappa}\theta^2 ,\n
\delta X^{\mu}&=&\delta_{SUSY}X^{\mu}+\delta_{\kappa}X^{\mu} ,
\eeqa
where we choose $\kappa^1$ and $\kappa^2$ as
\beqa
\kappa^1 &=& \frac{-\epsilon^1+\epsilon^2}{2} ,\n
\kappa^2 &=& \frac{\epsilon^1-\epsilon^2}{2} 
\eeqa
so that $\delta \theta^1 = \delta \theta^2$.
By introducing $\xi$ and $\epsilon$ as
\beqa
\xi &=& \frac{\epsilon^1+\epsilon^2}{2} ,\n
\epsilon &=& \frac{\epsilon^1-\epsilon^2}{2} ,
\eeqa
we can express the $N=2$ supersymmetry of $\tilde{S}_{GS}$  
as follows:
\beqa
\delta^{(1)}\psi &=& -\frac{1}{2\sqrt{-\frac{1}{2}\sigma^2}}
                      \sigma_{\mu\nu}\Gamma^{\mu\nu}\epsilon ,\n
\delta^{(1)} X^{\mu} &=& 4i\bar{\epsilon}\Gamma^{\mu}\psi ,
\label{NGsym1}
\eeqa
and
\beqa
\delta^{(2)}\psi &=& \xi ,\n
\delta^{(2)} X^{\mu} &=& 0 .
\label{NGsym2}
\eeqa

In order to rewrite $\tilde{S}_{GS}$ in the Schild gauge, we first introduce
the Poisson bracket as 
\beq
\{X,Y\} \equiv  \frac{1}{\sqrt{g}}\epsilon^{ab}\partial_a X \partial_b Y.
\eeq
Here $\sqrt{g}$ is a positive definite scalar density defined
on the world sheet which can be identified with $\sqrt{\det(g_{ab})}$ made from  
a worldsheet metric $g_{ab}$.
Using this, we write a Schild-type action \cite{Schild} as 
\beq
S_{Schild}=\int d^2\sigma [\sqrt{g}\alpha( 
\frac{1}{4}\{X^{\mu},X^{\nu}\}^2
-\frac{i}{2}\bar{\psi}\Gamma^{\mu}\{X^{\mu},\psi\})
+\beta \sqrt{g}].
\label{SSchild}
\eeq
As was shown by Schild some time ago, 
this action is equivalent to the original
action $\tilde{S}_{GS}$ (\ref{SGStilde})  
at least classically. Indeed, from the equation of motion
for $\sqrt{g}$ in (\ref{SSchild})
\beq
-\frac{1}{4}\alpha \frac{1}{(\sqrt{g})^2}
(\epsilon^{ab}\partial_a X^{\mu} \partial_b X^{\nu})^2+\beta =0 ,
\eeq
we obtain
\beq
\sqrt{g}=\frac{1}{2}\sqrt{\frac{\alpha}{\beta}}
\sqrt{(\epsilon^{ab}\partial_a X^{\mu} \partial_b X^{\nu})^2}.
\eeq
Then by substituting it into (\ref{SSchild}), we obtain
\beq
\int d^2 \sigma(\sqrt{\alpha\beta}
\sqrt{(\epsilon^{ab}\partial_a X^{\mu} \partial_b X^{\nu})^2}
-\frac{i}{2}\alpha \epsilon^{ab} \partial_a X^{\mu}
\bar{\psi}\Gamma^{\mu}\partial_b \psi),
\eeq
which is equivalent to $\tilde{S}_{GS}$ up to a normalization of
$\psi$, which we can adjust freely.
The $N=2$ supersymmetry (\ref{NGsym1}) and (\ref{NGsym2}) possessed 
by $\tilde{S}_{GS}$ manifests itself in  $S_{Schild}$ as 
\beqa
\delta^{(1)}\psi &=& -\frac{1}{2}
                      \sigma_{\mu\nu}\Gamma^{\mu\nu}\epsilon ,\n
\delta^{(1)} X^{\mu} &=& i\bar{\epsilon}\Gamma^{\mu}\psi ,
\label{SSchildsym1}
\eeqa
and
\beqa
\delta^{(2)}\psi &=& \xi ,\n
\delta^{(2)} X^{\mu} &=& 0 .
\label{SSchildsym2}
\eeqa
Although it is not clear to what extent the theory in this gauge is 
controllable as
a full quantum theory, at least formally the quantum theory is expressed
as a path integral over $X^{\mu}$ and $\sqrt{g}$:
\beq
{\cal Z} = \int\, {\cal D} \sqrt{g}\, {\cal D} X\, {\cal D} \psi \:e^{-S_{Schild}} .
\label{quantumSchild}
\eeq
Instead of trying to give a rigorous definition of (\ref{quantumSchild}), 
we first try to construct a matrix model which is equivalent to
(\ref{quantumSchild}) at least in the classical limit and interpret
its path integral as the definition of (\ref{quantumSchild}).

In the following, we show that the system defined by (\ref{quantumSchild})
can be regarded as a sort of the classical limit of the system defined by
\beq
Z= \sum_{n=0}^{\infty} \int dA \,d\psi \: e^{-S}
\label{quantumS}
\eeq
and
\beq
S=\alpha(- \frac{1}{4}Tr[A_{\mu},A_{\nu}]^2
            -\frac{1}{2}Tr(\bar{\psi}
           \Gamma^{\mu}[A_{\mu},\psi]))+\beta Tr \mbox{\bf $1$} .
\label{S}
\eeq
Here $A_{\mu}$ and $\psi$ are bosonic and fermionic 
$n \times n$ hermitian matrices respectively.
If large values of $n$ dominate in (\ref{quantumS}) and
the dominant distributions of eigenvalues for $A_{\mu}$ and $\psi$ are smooth
enough, we expect that the commutator and the trace can be replaced 
with the Poisson bracket
and the integration, respectively:  
\beqa
-i[\;,\;] &\Rightarrow& \{\;,\;\} ,\n
Tr &\Rightarrow& \int d^2 \sigma \sqrt{g} .
\label{correspondence}
\eeqa
This is the same as the ordinary correspondence between the quantum and
classical mechanics. As is well known, the basic properties of the commutator 
and the trace 
\beqa
Tr[X,Y] &=& 0 ,\n
Tr(X[Y,Z]) &=& Tr(Z[X,Y]) ,
\eeqa
are preserved after taking the classical limit: 
\beqa
\int d^2\sigma \sqrt{g}\{X,Y\} &=& 0 ,\n
\int d^2\sigma \sqrt{g}X\{Y,Z\}&=& \int d^2\sigma \sqrt{g}Z\{X,Y\} .
\eeqa
Now it is obvious that $S$ (\ref{S}) becomes $S_{Schild}$ (\ref{SSchild})
after this replacement \cite{Bars}. 
We also note that the sum over $n$ in (\ref{quantumS})
corresponds to the path integration over $\sqrt{g}$ in (\ref{quantumSchild}).
Furthermore we can easily check that
the $N=2$ supersymmetry (\ref{SSchildsym1}) and (\ref{SSchildsym2}) is
directly translated into
the symmetry of $S$ as
\beqa
\delta^{(1)}\psi &=& \frac{i}{2}
                     [A_{\mu},A_{\nu}]\Gamma^{\mu\nu}\epsilon ,\n
\delta^{(1)} A_{\mu} &=& i\bar{\epsilon}\Gamma_{\mu}\psi ,
\label{Ssym1}
\eeqa
and
\beqa
\delta^{(2)}\psi &=& \xi ,\n
\delta^{(2)} A_{\mu} &=& 0.
\label{Ssym2}
\eeqa

At this stage we can not claim that (\ref{quantumS}) gives a totally consistent
theory that satisfies the unitarity, causality and so on. However we will see
in section 4 that (\ref{quantumS}) can be regarded as an effective theory of
a more complete theory, that is, the large-N reduced model of super Yang-Mills 
theory. Here we simply point out that the form of $S$ in (\ref{S}) 
can be regarded
as a naive zero volume limit of ten-dimensional super Yang-Mills theory, 
except for the term proportional to $\beta$.
Then it is obvious that the symmetry (\ref{Ssym1}) is nothing but 
the zero volume limit of 
$N=1$ supersymmetry of the super Yang-Mills theory.  
To verify that the symmetries (\ref{Ssym1}) and (\ref{Ssym2}) indeed form 
the $N=2$ supersymmetry, we examine the commutators of these transformations
below. First of all we note that the action (\ref{S}) 
possesses the zero volume version of the gauge symmetry,
\beqa
\delta_{gauge} A_{\mu} &=& i[A_{\mu},\alpha], \n
\delta_{gauge} \psi    &=& i[\psi,\alpha].
\eeqa
Up to the above gauge symmetry 
and the equation of motion for $\psi$, we have the following 
commutation relations:
\beqa
(\delta^{(1)}_{\epsilon^1}\delta^{(1)}_{\epsilon^2}
    -\delta^{(1)}_{\epsilon^2}\delta^{(1)}_{\epsilon^1})\psi   &=&0 ,\n
(\delta^{(1)}_{\epsilon^1}\delta^{(1)}_{\epsilon^2}
    -\delta^{(1)}_{\epsilon^2}\delta^{(1)}_{\epsilon^1})A_{\mu}&=&0 .
\eeqa
We can also easily check the following commutators:
\beqa
(\delta^{(1)}_{\epsilon}\delta^{(2)}_{\xi}
    -\delta^{(2)}_{\xi}\delta^{(1)}_{\epsilon})\psi   &=&0 ,\n
(\delta^{(1)}_{\epsilon}\delta^{(2)}_{\xi}
    -\delta^{(2)}_{\xi}\delta^{(1)}_{\epsilon})A_{\mu}&=&
                     i\bar{\epsilon}\Gamma_{\mu}\xi ,  
\eeqa 
and
\beqa
(\delta^{(2)}_{\xi^1}\delta^{(2)}_{\xi^2}
    -\delta^{(2)}_{\xi^2}\delta^{(2)}_{\xi^1})\psi   &=&0 ,\n
(\delta^{(2)}_{\xi^1}\delta^{(2)}_{\xi^2}
    -\delta^{(2)}_{\xi^2}\delta^{(2)}_{\xi^1})A_{\mu}&=& 
                      0 .  
\eeqa 
If we take a linear combination of $\delta^{(1)}$ and $\delta^{(2)}$ as
\beqa 
\tilde{\delta}_{(1)}&=&\delta^{(1)}+\delta^{(2)}, \n
\tilde{\delta}_{(2)}&=&i(\delta^{(1)}-\delta^{(2)}),
\label{redefN=2}
\eeqa
we obtain the $N=2$ supersymmetry algebra,
\beqa
(\tilde{\delta}^{(1)}_{\epsilon}\tilde{\delta}^{(1)}_{\xi}
    -\tilde{\delta}^{(1)}_{\xi}\tilde{\delta}^{(1)}_{\epsilon})\psi   &=&0 ,\n
(\tilde{\delta}^{(1)}_{\epsilon}\tilde{\delta}^{(1)}_{\xi}
    -\tilde{\delta}^{(1)}_{\xi}\tilde{\delta}^{(1)}_{\epsilon})A_{\mu}&=&
                                 2i\bar{\epsilon}\Gamma_{\mu}\xi , \n
(\tilde{\delta}^{(2)}_{\epsilon}\tilde{\delta}^{(2)}_{\xi}
    -\tilde{\delta}^{(2)}_{\xi}\tilde{\delta}^{(2)}_{\epsilon})\psi   &=&0 ,\n
(\tilde{\delta}^{(2)}_{\epsilon}\tilde{\delta}^{(2)}_{\xi}
    -\tilde{\delta}^{(2)}_{\xi}\tilde{\delta}^{(2)}_{\epsilon})A_{\mu}&=&
                                 2i\bar{\epsilon}\Gamma_{\mu}\xi , \n
(\tilde{\delta}^{(1)}_{\epsilon}\tilde{\delta}^{(2)}_{\xi}
    -\tilde{\delta}^{(2)}_{\xi}\tilde{\delta}^{(1)}_{\epsilon})\psi   &=&0 ,\n
(\tilde{\delta}^{(1)}_{\epsilon}\tilde{\delta}^{(2)}_{\xi}
    -\tilde{\delta}^{(2)}_{\xi}\tilde{\delta}^{(1)}_{\epsilon})A_{\mu}&=&0 .
\label{Nequal2}
\eeqa

The system we consider here has the manifest Lorentz invariance 
and is probably unitary. 
Therefore the global $N=2$ supersymmetry (\ref{Nequal2}) is extended to
the local one inevitably since the $N=1$ supersymmetry is maximal
in ten-dimensional gauge theory. If the theory admits massless spectrum, 
it contains
gravitons. Furthermore as we discuss just below, 
the path integral (\ref{quantumS}) automatically includes not only one-string
states but also multi-string states.
When the matrices $A_{\mu}$'s and $\psi$ are block-diagonal, the action is 
decomposed into the sum of traces for each of the blocks. Each trace results
in the string action in the Schild gauge in the classical limit.
Then the trace in (\ref{S}) corresponds to the integrals
over the disconnected worldsheets. Namely, each block represents a string and
the theory defined by (\ref{quantumS}) include multi-strings states.
As we will see in the next section, the path integral of the off-diagonal-blocks
generate interactions between the diagonal-blocks. 
Because of these facts, we expect that the theory (\ref{quantumS}) after
a slight modification discussed in section 4 possibly gives 
a constructive definition of the string theory.


\vspace{1cm}

\section{One-loop quantum corrections around classical solutions}
\setcounter{equation}{0}
\subsection{Classical static D-string solutions}
\hspace*{\parindent}
In this section, we consider the typical classical solutions of (\ref{S})
which represent infinitely long static D-strings \cite{Polchinski}.
When $\psi=0$, the equation of motion of (\ref{SSchild}) is
\beq
\{X^{\mu},\{X^{\mu},X^{\nu}\}\}=0.
\label{SSchildEOM}
\eeq
Corresponding to this, the equation of motion of (\ref{S}) is
\beq
[A_{\mu},[A_{\mu},A_{\nu}]]=0.
\label{SEOM}
\eeq
We can easily construct a solution of (\ref{SSchildEOM}), which represents
a static D-string extending straightly in the $X^1$ direction:
\beqa
X^0&=&T \tau ,\n
X^1&=&\frac{L}{2\pi}\sigma ,\n
\mbox{other }X^{\mu} \mbox{'s}&=& 0,
\label{Schildstaticsolution}
\eeqa
where $T$ and $L$ are large enough compactification radii and
\beqa
0 \leq &\tau& \leq 1, \n
0 \leq &\sigma& \leq 2\pi. 
\eeqa
Considering the relation between the commutator and the Poisson bracket,
we obtain a solution of (\ref{SEOM}) corresponding to the above one as follows:
\beqa
A_0 &=& \frac{T}{\sqrt{2\pi n}}q \equiv p_0 ,\n
A_1 &=& \frac{L}{\sqrt{2\pi n}}p \equiv p_1 ,\n
\mbox{other }A_{\mu} \mbox{'s}&=& 0,
\label{Sstaticsolution}
\eeqa
where $T$ and $L$ are large enough compactification radii, and $q$ and $p$ are 
$n \times n$ hermitian matrices having the following commutation relation and
the eigenvalue distributions: 
\beq
[q,p]=i,\\
\label{qpcommutator}
\eeq
and
\beqa
&~&0 \leq q \leq {\sqrt{2\pi n}},\n
&~&0 \leq p \leq {\sqrt{2\pi n}}.
\label{qp}
\eeqa
Strictly speaking such $p$ and $q$ do not exist for finite values of $n$.
For large values of $n$, however, we expect that (\ref{qpcommutator}) can be
approximately satisfied, because it is nothing but the canonical commutation
relation. As is well-known in the correspondence between the classical and
quantum mechanics, the total area of the $p-q$ phase space is equal to $2\pi$
multiplied by the dimension of the representation. In this sense (\ref{qp})
indicates that $p$ and $q$ are $n \times n$ matrices.

We can also construct classical solutions corresponding to two static D-strings 
in the same way. As was mentioned at the end of the previous section, 
we can obtain
these solutions by considering $A_{\mu}$'s composed of 
two diagonal-blocks. 
First, we consider two parallel D-strings, which extend infinitely in the $X^1$
direction and separated by distance $b$ in the $X^2$ direction.
The solution of (\ref{SSchildEOM}) representing this situation is given by
\beq
\left\{
\begin{array}{rcl}
X^{(1)0}&=&T \tau^{(1)} ,\\
X^{(1)1}&=&\frac{L}{2\pi}\sigma^{(1)} ,\\
X^{(1)2}&=&\frac{b}{2}, \\
\mbox{other }X^{(1)\mu} \mbox{'s}&=& 0, 
\end{array} \right. \:
\left\{
\begin{array}{rcl}
X^{(2)0}&=&T \tau^{(2)} ,\\
X^{(2)1}&=&\frac{L}{2\pi}\sigma^{(2)} ,\\
X^{(2)2}&=&-\frac{b}{2}, \\
\mbox{other }X^{(2)\mu} \mbox{'s}&=& 0 .
\end{array} \right.
\eeq
By identifying the first block with the first D-string and the second block
with the second D-string,
we obtain the solution of (\ref{SEOM}) corresponding to the above one 
\beqa
A_{0}&=&\left(\begin{array}{cc}
                    \frac{T}{\sqrt{2\pi n}}q  &     0        \\      
                           0          &  \frac{T}{\sqrt{2\pi n}}q^{'}  \\   
                                \end{array} \right) \equiv p_0 ,\n       
A_{1}&=&\left(\begin{array}{cc}
                    \frac{L}{\sqrt{2\pi n}}p  &     0         \\      
                           0            &  \frac{L}{\sqrt{2\pi n}}p^{'}  \\   
                                \end{array} \right) \equiv p_1 ,\n       
A_{2}&=&\left(\begin{array}{cc}
                          \frac{b}{2} &     0        \\      
                                0     &  -\frac{b}{2} \\   
                                \end{array} \right) \equiv p_2 ,\n       
\mbox{other } A_{\mu} \mbox{'s} &=& 0, 
\label{paralell}
\eeqa
where $q^{'}$ and $p^{'}$ satisfy the same properties as (\ref{qpcommutator})
and (\ref{qp}).
Applying a unitary transformation, we can set $q^{'}$ equal to $q$ 
and $p^{'}$ equal to $p$. 
Similarly for the solution of (\ref{SSchildEOM}) representing two anti-parallel
D-strings
\beq
\left\{
\begin{array}{rcl}
X^{(1)0}&=&T \tau^{(1)} ,\\
X^{(1)1}&=&\frac{L}{2\pi}\sigma^{(1)} ,\\
X^{(1)2}&=&\frac{b}{2}, \\
\mbox{other }X^{(1)\mu} \mbox{'s}&=& 0, 
\end{array} \right. \:
\left\{
\begin{array}{rcl}
X^{(2)0}&=&T \tau^{(2)} ,\\
X^{(2)1}&=&-\frac{L}{2\pi}\sigma^{(2)} ,\\
X^{(2)2}&=&-\frac{b}{2}, \\
\mbox{other }X^{(2)\mu} \mbox{'s}&=& 0 ,
\end{array} \right.
\eeq
we obtain the corresponding solution of (\ref{SEOM})
\beqa
A_{0}&=&\left(\begin{array}{cc}
                    \frac{T}{\sqrt{2\pi n}}q  &     0        \\      
                           0          &  \frac{T}{\sqrt{2\pi n}}q  \\   
                                \end{array} \right) \equiv p_0 ,\n       
A_{1}&=&\left(\begin{array}{cc}
                    \frac{L}{\sqrt{2\pi n}}p  &     0         \\      
                           0            &  -\frac{L}{\sqrt{2\pi n}}p  \\   
                                \end{array} \right) \equiv p_1 ,\n       
A_{2}&=&\left(\begin{array}{cc}
                          \frac{b}{2} &     0        \\      
                                0     &  -\frac{b}{2} \\   
                                \end{array} \right) \equiv p_2 ,\n       
\mbox{other } A_{\mu} \mbox{'s} &=& 0.
\label{antiparalell}
\eeqa
Finally, we consider a general configurations of two static straight D-strings 
described by 
\beq
\left\{
\begin{array}{rcl}
X^{(1)0}&=&T \tau^{(1)} ,\\
X^{(1)1}&=&\frac{L}{2\pi}\sigma^{(1)} ,\\
X^{(1)2}&=&0 ,\\
X^{(1)3}&=&\frac{b}{2} ,\\
\mbox{other }X^{(1)\mu} \mbox{'s}&=& 0, 
\end{array} \right. \:
\left\{
\begin{array}{rcl}
X^{(2)0}&=&T \tau^{(2)} ,\\
X^{(2)1}&=&\frac{L}{2\pi}\sigma^{(2)} \, \cos \theta ,\\
X^{(2)2}&=&\frac{L}{2\pi}\sigma^{(2)} \, \sin \theta ,\\
X^{(2)3}&=&-\frac{b}{2}, \\
\mbox{other }X^{(2)\mu} \mbox{'s}&=& 0 ,
\end{array} \right.
\eeq
where $b$ is the minimum distance between the  two D-strings . The solution
of (\ref{SEOM}) corresponding to the above one is
\beqa
A_0&=&\left(\begin{array}{cc}
                    \frac{T}{\sqrt{2\pi n}} q &           0              \\      
                               0             &  \frac{T}{\sqrt{2\pi n}} q     
                       \end{array} \right) \equiv p_0 ,\n
A_1&=&\left(\begin{array}{cc}
         \frac{L}{\sqrt{2\pi n}} p &           0              \\      
                    0             &  \frac{L}{\sqrt{2\pi n}} p \cos \theta    
                       \end{array} \right) \equiv p_1 ,\n
A_2&=&\left(\begin{array}{cc}
                        0          &           0              \\      
                        0          &  \frac{L}{\sqrt{2\pi n}} p \sin \theta   
                       \end{array} \right) \equiv p_2 ,\n
A_3&=&\left(\begin{array}{cc}
                      \frac{b}{2}   &           0              \\      
                           0        &      -\frac{b}{2}    
                       \end{array} \right) \equiv p_3 ,\n
\mbox{other } A_{\mu} \mbox{'s} &=& 0 .
\label{generalconfiguration}
\eeqa

\subsection{One-loop effective action and stability of BPS-saturated states}
\hspace*{\parindent}
In appendix, we calculate the one-loop effective action $W$ 
around a general background $A_{\mu}=p_{\mu}$. The result is
\beq
ReW  = \frac{1}{2}Tr \log(P_{\lambda}^2 \delta_{\mu\nu}-2iF_{\mu\nu})
      -\frac{1}{4}Tr \log((P_{\lambda}^2+\frac{i}{2}F_{\mu\nu}\Gamma^{\mu\nu})
      (\frac{1+\Gamma_{11}}{2}))-Tr \log(P_{\lambda}^2).
\label{oneloopeffpot}
\eeq
Here $P_{\mu}$ and $F_{\mu\nu}$ are operators acting on the space of matrices as
\beqa
        P_{\mu}X & = & [p_{\mu},X] ,\n
        F_{\mu\nu}X & = & \left[ f_{\mu\nu},X \right],
\label{adjointoperator}
\eeqa
where $f_{\mu \nu}=i[p_{\mu},p_{\nu}]$. 
In (\ref{oneloopeffpot}), each of the three terms corresponds to the 
contributions 
from the bosons $A_{\mu}$, the fermions $\psi$ 
and the Fadeev-Popov ghosts, respectively.

The cases in which $f_{\mu \nu}=c-number\equiv c_{\mu\nu}$ 
have a special meaning.
These correspond to BPS-saturated backgrounds \cite{WO}. Indeed, by setting
$\xi$ equal to $\pm \frac{1}{2}c_{\mu\nu} \Gamma^{\mu\nu}\epsilon$ 
in the $N=2$ supersymmetry (\ref{Ssym1}) and (\ref{Ssym2}),
we obtain the relations
\beqa
(\delta^{(1)}\mp \delta^{(2)}) \psi &=&0 ,\n
(\delta^{(1)}\mp \delta^{(2)}) A_{\mu}&=& 0.
\eeqa
Namely, half of the supersymmetry is preserved in these backgrounds.
Since $F_{\mu\nu}=0$ in these cases, we have
\beq
ReW=(\frac{1}{2}\cdot 10 -\frac{1}{4} \cdot 16 -1)Tr \log(P_{\lambda}^2)=0 ,
\eeq
and
\beq
ImW=0,
\eeq
as is shown in appendix,
which means the one-loop quantum corrections vanish due to the supersymmetry. 
This is consistent with the well-known fact that the BPS-saturated states have no
corrections and are stable. One of the simplest examples of the BPS-saturated 
states is the solution (\ref{Sstaticsolution}). Indeed, for this solution,
\beqa
f_{01}=i[p_{0},p_{1}]&=&-\frac{TL}{2\pi n} ,\n
\mbox{other } f_{\mu \nu} \mbox{'s}&=&0 .
\eeqa
We see that an infinitely long static D-string exists stably.

More specifically we may consider the cases in which $c_{\mu\nu}=0$, that is, 
$[p_{\mu},p_{\nu}]=0$. Then we can diagonalize 
all $p_{\mu}$'s simultaneously. These are classical minima of the action and
are often called moduli in supersymmetric gauge theories.
The one-loop effective 
action also vanishes in these cases, and there are no interactions
between the eigenvalues. Namely, the moduli space is stable even 
quantum mechanically,
which is a manifestation of the non-renormalization theorem in supersymmetric
theories. The eigenvalues tend to spread randomly and their
distribution is likely to be uniform, which means that
$U(1)^{d}$ symmetry is marginally preserved. Note that if we had
no contributions from fermions, 
there would be attractive logarithmic interactions
between the eigenvalues and they would concentrate on one value. This is
the well-known $U(1)^d$ symmetry breaking in the weak coupling region of
non-supersymmetric gauge theories. 


\subsection{Interactions between two static D-strings}
\hspace*{\parindent}
First we consider the solution (\ref{paralell}) representing
the two parallel static D-strings. Since in this case
\beqa
f_{01}=i[p_{0},p_{1}]&=&-\frac{TL}{2\pi n} ,\n
\mbox{other } f_{\mu \nu} \mbox{'s}&=&0,
\eeqa
this solution represents a BPS-saturated state and the one-loop effective
action is equal to zero.  
We obtain a consistent picture that there is no force 
between two parallel D-strings
due to the cancellation of the gravitational force with the force mediated by 
the 
anti-symmetric tensor field. This implies the possibility of superposing
BPS-saturated states.

As a preparation of the calculations of the quantum corrections 
to the solutions (\ref{antiparalell})
and (\ref{generalconfiguration}), we evaluate the one-loop effective action
(\ref{oneloopeffpot}) when $[p_{\mu},f_{\nu\lambda}]=c-number$.  
In these cases $P_{\lambda}^2$ and $F_{\mu\nu}$ commute with each other and are
simultaneously diagonalizable.
We set $F_{\mu\nu}$ in the standard form
\beq
F_{\mu\nu}=\left(\begin{array}{ccccc}
                    0 & -a_1 &        &     &      \\      
                  a_1 &    0 &        &     &      \\   
                      &      & \ddots &     &       \\     
                      &      &        &   0 & -a_5  \\
                      &      &        & a_5 &    0     

                 \end{array} \right) ,
\eeq
and calculate each term of (\ref{oneloopeffpot}) as follows.
The first term is
\beq
Tr \log(P_{\lambda}^2 \delta_{\mu\nu}-2iF_{\mu\nu})=
\sum_{i=1}^{5} Tr \log((P_{\lambda}^2)^2-4a_i^2).
\label{firstterm}
\eeq
Considering that the eigenvalues of $\Gamma^{\mu\nu}$ are equal to $\pm i$ and
that we project the system to the space in which the eigenvalue of
$\Gamma_{11}=i\Gamma^1 \Gamma^2 \cdots \Gamma^{10}$ is equal to $1$, we
evaluate the second term of (\ref{oneloopeffpot}) as
\beq
Tr \log((P_{\lambda}^2+\frac{i}{2}F_{\mu\nu}\Gamma^{\mu\nu}))
      (\frac{1+\Gamma_{11}}{2}))
=\sum_{s_1,\ldots,s_5=\pm 1,s_1 \cdots s_5=1} 
       Tr \ log(P_{\lambda}^2-(a_1s_1+\cdots+a_5s_5)).
\label{secondterm}
\eeq
{}From (\ref{oneloopeffpot}), (\ref{firstterm}) and (\ref{secondterm}), we obtain
\beqa 
ReW & = & \frac{1}{2}\sum_{i=1}^{5} Tr log(1-\frac{4a_i^2}{(P_{\lambda}^2)^2}) \n
  &   & -\frac{1}{4}
          \sum_{s_1,\ldots,s_5=\pm 1,s_1 \cdots s_5=1} 
          Tr log(1-\frac{a_1s_1+\cdots+a_5s_5}{P_{\lambda}^2}).
\label{1loopeff}
\eeqa

We apply this expression to the cases we consider.
Since these cases correspond to case(2) in appendix, we have $ImW=0$.
For the solution (\ref{antiparalell}) corresponding to the two anti-parallel
static D-strings, we have
\beqa
f_{01}=i[p_{0},p_{1}]=-\frac{TL}{2\pi n}\otimes \sigma^3 ,\n
\mbox{other } f_{\mu \nu} \mbox{'s}=0 ,
\eeqa
and therefore
\beq
[p_{\mu},f_{\nu\lambda}]=0,
\eeq
where $\sigma^3$ is the third component of Pauli matrices.
As in (\ref{adjointoperator}) we define the adjoint operator $\Sigma^3$ 
corresponding to $1 \otimes \sigma^3$ as
\beq
\Sigma^3 X = [1 \otimes \sigma^3,X].  
\eeq
Then we have
\beq
[P_{0},P_1]=i\frac{TL}{2\pi n}\Sigma^3,
\label{adjcommutator}
\eeq
and
\beq
P_2=\frac{b}{2}\Sigma^3.
\eeq

The eigenvalues of $\Sigma^3$ are equal to
$0, 0, 2$ and $-2$. If the eigenvalue is equal to zero, $F_{\mu\nu}=0$
and there is no contribution to the one-loop effective action. If it is
equal to $\pm 2$, $P_{\lambda}^2$ in (\ref{1loopeff}) behaves like 
a harmonic oscillator because of the commutator (\ref{adjcommutator}), whose
eigenvalues are $4(\frac{TL}{2\pi n})(k+\frac{1}{2})+b^2$.
Each of these eigenvalues has $n$-fold degeneracy, 
because the operator $P_{\lambda}^2$ is acting on the space of $n \times n$
matrices. 
Therefore we can calculate (\ref{1loopeff}) in this case as
\beqa
W&=&-n \log \prod_{k=0}^{\infty}
      \frac{(1+\frac{1}{2k+1+\frac{{b^{'}}^2}{2}})^4 
            (1-\frac{1}{2k+1+\frac{{b^{'}}^2}{2}})^4}
           {(1+\frac{2}{2k+1+\frac{{b^{'}}^2}{2}}) 
            (1-\frac{2}{2k+1+\frac{{b^{'}}^2}{2}})} \n
 &=&-n \log\left[\left(\frac{{b^{'}}^2}{4}\right)^{-4}
       \frac{\frac{{b^{'}}^2}{4}+\frac{1}{2}}{\frac{{b^{'}}^2}{4}-\frac{1}{2}}
\left( \frac{\Gamma(\frac{{b^{'}}^2}{4}+\frac{1}{2})}
            {\Gamma(\frac{{b^{'}}^2}{4})}                                                                    \right)^8 \right],
\label{Wantiparalelloriginal}
\eeqa
where $b^{'}=\sqrt{\frac{2\pi n}{TL}}b$. This expression can be expanded
with respect to  $1/b^2$ as 
\beq
W=-8n(\frac{TL}{2\pi n})^3\frac{1}{b^6}+O(\frac{1}{b^8}).
\label{Wantiparalell}
\eeq
We see that there is an interaction which is mediated by massless particles 
between the two
anti-parallel D-strings, which should be twice as large as 
the gravitational interaction because of that of anti-symmetric tensor field.

Next, we evaluate the interaction between the two generally located
static D-strings described by (\ref{generalconfiguration}). 
If we introduce $p_1^{'}$ and $p_2^{'}$ as 
\beqa
p_1^{'}&=&\sin \frac{\theta}{2}\,p_1 -\cos\frac{\theta}{2} \,p_2 
        =\frac{L}{\sqrt{2\pi n}}p \sin\frac{\theta}{2}\otimes \sigma^3  ,\n
p_2^{'}&=&\cos \frac{\theta}{2}\,p_1 +\sin\frac{\theta}{2} \,p_2 
        =\frac{L}{\sqrt{2\pi n}}p \cos\frac{\theta}{2}\otimes 1_2 ,
\eeqa
the background field strength is expressed as
\beqa
f_{01}^{'}&=&i[p_0,p_1^{'}]
           =-\frac{TL}{2\pi n}\sin \frac{\theta}{2}\otimes \sigma^3  ,\n
f_{02}^{'}&=&i[p_0,p_2^{'}]
           =-\frac{TL}{2\pi n}\cos \frac{\theta}{2}\otimes  1_2 ,\n
f_{12}^{'}&=&0 ,
\eeqa
and hence
\beqa
F_{01}^{'}&=&-\frac{TL}{2\pi n}\sin \frac{\theta}{2}\Sigma^3 ,\n
\mbox{other }F_{\mu\nu}^{'} \mbox{'s} &=& 0
\eeqa
Therefore the calculation of the one-loop effective action can be reduced to 
that of the two anti-parallel D-strings case, except for the nontrivial
$P_2^{'}$, which commutes with the other $P_{\mu}^{'}$'s.
In this case the eigenvalues of $(P_{2}^{'})^2$ are distributed uniformly
from $0$ to $(\frac{L}{\sqrt{2\pi n}}\sqrt{2\pi n}\cos \frac{\theta}{2})^2$. 
Therefore
instead of the $n$-fold degeneracy in (\ref{Wantiparalelloriginal}), this time
we should integrate over the eigenvalues of $P_{2}^{'}$. Thus for large 
values of $b$ we have 
\beqa
W&=&-8\sqrt{\frac{n}{2\pi}}\int_0^{\sqrt{2\pi n}} dp \,
     \frac{(\frac{TL}{2\pi n}\sin \frac{\theta}{2})^3}
          {(b^2+(p_2^{'})^2)^3} \n
 &=&-\frac{3\pi}{2} n (\frac{TL}{2\pi n})^3 \frac{1}{L b^5}
                             \sin^2 \frac{\theta}{2} \tan\frac{\theta}{2} , 
\label{generalconfigW1}
\eeqa
where
\beq
p_2^{'}=\frac{L}{\sqrt{2\pi n}} p \cos \frac{\theta}{2}.
\eeq
In the following, we show that this result can be interpreted as
the exchange of anti-symmetric tensor field $B_{\mu\nu}$ and graviton 
$G_{\mu\nu}$ between the two D-strings. 
In the Lorentz-type gauge, the potential created by the first
D-string at distance $r$ is given in  the leading order
of $1/r$ by
\beqa
B_{01}&=& \frac{2{\alpha^{'}}^3}{r^6} ,\n
G_{00}&=& 1-\frac{{\alpha^{'}}^3}{r^6} ,\n
G_{11}&=& 1-\frac{{\alpha^{'}}^3}{r^6}.
\eeqa
On the other hand, the source for the anti-symmetric tensor
field and the energy momentum tensor carried by the second D-string are given by
\beq
J^{01}=\frac{1}{\alpha^{'}}\cos \theta ,
\eeq
and
\beqa
T^{00}&=&\frac{1}{\alpha^{'}} ,\n
T^{11}&=&\frac{1}{\alpha^{'}}\cos^2 \theta ,
\eeqa
respectively.
Therefore a fraction $Ld\sigma$ of the second D-string located 
at distance $r$ from the first one possesses the potential energy
\beq
2{\alpha^{'}}^2\frac{\cos \theta}{r^6}-{\alpha^{'}}^2\frac{1+\cos^2\theta}{r^6}
=-4{\alpha^{'}}^2\frac{\sin^4 \frac{\theta}{2}}{r^6} ,
\eeq
and the effective action is evaluated as
\beq
-TL\int_0^{2\pi} d\sigma \frac{{\alpha^{'}}^2\sin^4 \frac{\theta}{2}}
                         {(b^2+L^2\sigma^2\sin^2 \theta)^3}
=-n \frac{TL}{2\pi n}
    \frac{{\alpha^{'}}^2}{b^5 L} \sin^2 \frac{\theta}{2} \tan\frac{\theta}{2}.
\eeq
As we will see in section 3.4, $TL/2\pi n$ is identified with $\alpha^{'}$,
and the above expression agrees with (\ref{generalconfigW1}).

\subsection{Interactions between diagonal blocks and cluster property}
\hspace*{\parindent}
In this subsection, we calculate the one-loop effective action between
diagonal blocks 
in order to identify the interactions included in the effective action with those
of string theory.
We consider backgrounds having a block-diagonal form: 
\beq
A_{\mu}=p_{\mu}=\left(\begin{array}{cccc}
                    p_{\mu}^{(1)}  &                &               &   \\      
                                   & p_{\mu}^{(2)}  &               &   \\   
                                   &                & p_{\mu}^{(3)} &   \\
                                   &                &               &  \ddots
                 \end{array} \right) ,
\label{blockdiagonal}
\eeq
where $p_{\mu}^{(i)}$ $(i=1, 2, \cdots)$ is a $n_i\times n_i$ matrix.
We may regard each $p_{\mu}^{(i)}$ as a D-object 
occupying some region of space-time.
Here we use a term D-object to represent D-instantons, D-strings, D-3 branes,
$\ldots$ and their mixtures. 
We decompose $p_{\mu}^{(i)}$ as
\beqa
p_{\mu}^{(i)}&=&d_{\mu}^{(i)}1_{n_i}+\tilde{p}_{\mu}^{(i)} ,\n 
Tr\tilde{p}_{\mu}^{(i)}&=& 0 ,
\label{cm}
\eeqa
where $d_{\mu}^{(i)}$ is a real number representing the center of mass 
coordinate of the $i$-th block.
Here we assume that the blocks are separated far enough from each other, that is,
for all $i$ and $j$'s, $(d_{\mu}^{(i)}-d_{\mu}^{(j)})^2$'s are large. 

We first introduce some notations. 
We denote the $(i,j)$ block of a matrix $X$ as $X^{(i,j)}$. It is clear that
$P_{\mu}$ defined in (\ref{adjointoperator}) operates on each $X^{(i,j)}$
independently. In fact we have
\beq
(P_{\mu}X)^{(i,j)}=(d_{\mu}^{(i)}-d_{\mu}^{(j)}) X^{(i,j)}
                      +\tilde{p}_{\mu}^{(i)}X^{(i,j)}
                                    -X^{(i,j)}\tilde{p}_{\mu}^{(j)}. 
\eeq
We further simplify this equation by introducing notations such as
\beqa
d_{\mu}^{(i,j)}X^{(i,j)}&=&(d_{\mu}^{(i)}-d_{\mu}^{(j)}) X^{(i,j)} ,\n
P_{L\mu}^{(i,j)}X^{(i,j)}&=&\tilde{p}_{\mu}^{(i)}X^{(i,j)} ,\n
P_{R\mu}^{(i,j)}X^{(i,j)}&=&-X^{(i,j)}\tilde{p}_{\mu}^{(j)} .
\eeqa
Note that $d_{\mu}^{(i,j)}$,$P_{L\mu}^{(i,j)}$ and $P_{R\mu}^{(i,j)}$ 
commute each other, and
the operation of $P_{\mu}$ on $X^{(i,j)}$ is expressed as
\beq
P_{\mu}X^{(i,j)}=(d_{\mu}^{(i,j)}+P_{L\mu}^{(i,j)}
                              +P_{R\mu}^{(i,j)})X^{(i,j)}.
\label{lrdecomposition}
\eeq
We similarly simplify the form of $F_{\mu\nu}$.
The background field strength is evaluated as
\beqa
f_{\mu\nu}&=&\left(\begin{array}{ccc}
                    i[p_{\mu}^{(1)},p_{\nu}^{(1)}] &  &         \\     
                              & i[p_{\mu}^{(2)},p_{\nu}^{(2)}]  &  \\        
                                   &                & \ddots     
                 \end{array} \right) \n
          &=&\left(\begin{array}{ccc}
            \tilde{f}_{\mu\nu}^{(1)}  &                   &             \\      
                     & \tilde{f}_{\mu\nu}^{(2)}  &              \\   
                                   &                      & \ddots     
                 \end{array} \right) ,
\eeqa
and we can also decompose $F_{\mu\nu}$ 
in the same way as (\ref{lrdecomposition}):
\beq
(F_{\mu\nu}X)^{(i,j)}=(F_{L\mu\nu}^{(i,j)}
                          +F_{R\mu\nu}^{(i,j)})X^{(i,j)},
\label{lrdecompositionF}
\eeq
where
\beqa
F_{L\mu\nu}^{(i,j)}X^{(i,j)}&=&\tilde{f}_{\mu\nu}^{(i)}X^{(i,j)} ,\n
F_{R\mu\nu}^{(i,j)}X^{(i,j)}&=&-X^{(i,j)}\tilde{f}_{\mu\nu}^{(j)}.
\eeqa
Since the left and right multiplication are totally independent, we have
\beq
Tr \, O =\sum_{i,j=1}^{n} Tr O_L^{(i,j)} Tr O_R^{(i,j)},
\label{ijtrace}
\eeq
for operators consisting of $P_{\mu}$ and $F_{\mu\nu}$.

Now we expand the general expression of the one-loop effective action
(\ref{oneloopeffpot}) with respect to the inverse power of $d_{\mu}^{(i,j)}$'s.
We can take traces of the $\gamma$ matrices after expanding the logarithm in
(\ref{oneloopeffpot}).
Due to the supersymmetry, contributions of bosons and fermions cancel 
each other to the third order in $F_{\mu\nu}$, and we have, 
\beqa
W    &=&-Tr\left(\frac{1}{P^2}F_{\mu\nu} \frac{1}{P^2}F_{\nu\lambda}
             \frac{1}{P^2}F_{\lambda\rho}\frac{1}{P^2}F_{\rho\mu} \right) \n
       &~& -2Tr\left(\frac{1}{P^2}F_{\mu\nu} \frac{1}{P^2}F_{\lambda\rho}
             \frac{1}{P^2}F_{\mu\rho}\frac{1}{P^2}F_{\lambda\nu} \right) \n
       &~&+\frac{1}{2}Tr\left(\frac{1}{P^2}F_{\mu\nu} \frac{1}{P^2}F_{\mu\nu}
             \frac{1}{P^2}F_{\lambda\rho}\frac{1}{P^2}F_{\lambda\rho} \right) \n
      &~&+\frac{1}{4}Tr\left(\frac{1}{P^2}F_{\mu\nu} \frac{1}{P^2}F_{\lambda\rho}
             \frac{1}{P^2}F_{\mu\nu}\frac{1}{P^2}F_{\lambda\rho} \right)
       +O((F_{\mu\nu})^5).
\label{Wexpansion}
\eeqa
Since as in (\ref{lrdecomposition}) and (\ref{lrdecompositionF}) $P_{\mu}$
and $F_{\mu\nu}$ act on the $(i,j)$ blocks independently, the one-loop effective
action $W$ is expressed as the sum of contributions of the $(i,j)$ blocks
$W^{(i,j)}$. Therefore we may consider $W^{(i,j)}$
as the interaction between the i-th and j-th blocks. Using (\ref{ijtrace})
and (\ref{Wexpansion}) we can easily evaluate $W^{(i,j)}$
to the leading order of $1/\sqrt{(d^{(i)}-d^{(j)})^2}$ as
\beqa
W^{(i,j)}&=&\frac{1}{(d^{(i)}-d^{(j)})^8} \n
         & &( -Tr^{(i,j)}\left(F_{\mu\nu}F_{\nu\lambda}
                            F_{\lambda\rho}F_{\rho\mu} \right) 
                  -2Tr^{(i,j)}\left(F_{\mu\nu}F_{\lambda\rho}
                             F_{\mu\rho}F_{\lambda\nu} \right) \n
         & &+\frac{1}{2}Tr^{(i,j)}\left(F_{\mu\nu}F_{\mu\nu}
                                      F_{\lambda\rho}F_{\lambda\rho} \right) 
                +\frac{1}{4}Tr^{(i,j)}\left(F_{\mu\nu}F_{\lambda\rho}
                                     F_{\mu\nu}F_{\lambda\rho} \right))\n 
         & &  +O((1/(d^{(i)}-d^{(j)})^9)\n
         &=&\frac{1}{4(d^{(i)}-d^{(j)})^8}\n
         & &(-4n_jTr(\tilde{f}^{(i)}_{\mu\nu}\tilde{f}^{(i)}_{\nu\lambda}
                  \tilde{f}^{(i)}_{\lambda\rho}\tilde{f}^{(i)}_{\rho\mu})
             -8n_jTr(\tilde{f}^{(i)}_{\mu\nu}\tilde{f}^{(i)}_{\lambda\rho}
                  \tilde{f}^{(i)}_{\mu\rho}\tilde{f}^{(i)}_{\lambda\nu})\n
         & &+2n_jTr(\tilde{f}^{(i)}_{\mu\nu}\tilde{f}^{(i)}_{\mu\nu}
                  \tilde{f}^{(i)}_{\lambda\rho}\tilde{f}^{(i)}_{\lambda\rho})
            +n_jTr(\tilde{f}^{(i)}_{\mu\nu}\tilde{f}^{(i)}_{\lambda\rho}
                  \tilde{f}^{(i)}_{\mu\nu}\tilde{f}^{(i)}_{\lambda\rho})\n
         & &-4n_iTr(\tilde{f}^{(j)}_{\mu\nu}\tilde{f}^{(j)}_{\nu\lambda}
                  \tilde{f}^{(j)}_{\lambda\rho}\tilde{f}^{(j)}_{\rho\mu})
             -8n_iTr(\tilde{f}^{(j)}_{\mu\nu}\tilde{f}^{(j)}_{\lambda\rho}
                  \tilde{f}^{(j)}_{\mu\rho}\tilde{f}^{(j)}_{\lambda\nu})\n
         & &+2n_iTr(\tilde{f}^{(j)}_{\mu\nu}\tilde{f}^{(j)}_{\mu\nu}
                  \tilde{f}^{(j)}_{\lambda\rho}\tilde{f}^{(j)}_{\lambda\rho})
            +n_iTr(\tilde{f}^{(j)}_{\mu\nu}\tilde{f}^{(j)}_{\lambda\rho}
                  \tilde{f}^{(j)}_{\mu\nu}\tilde{f}^{(j)}_{\lambda\rho})\n
         & &-48Tr(\tilde{f}^{(i)}_{\mu\nu}\tilde{f}^{(i)}_{\nu\lambda})
              Tr(\tilde{f}^{(j)}_{\mu\rho}\tilde{f}^{(j)}_{\rho\lambda})
            +6Tr(\tilde{f}^{(i)}_{\mu\nu}\tilde{f}^{(i)}_{\mu\nu})
            Tr(\tilde{f}^{(j)}_{\lambda\rho}\tilde{f}^{(j)}_{\lambda\rho}))\n 
         & &  +O((1/(d^{(i)}-d{(j)})^9).
\label{blockinteraction}
\eeqa
By observing the tensor structures of the last two terms in
(\ref{blockinteraction}), we find the exchanges of massless particles
corresponding to graviton and a scalar.

As we have seen, the interactions between two
blocks are weaker than or equal to $1/r^8$, where $r$ is the distance
between two centers of mass. Therefore if D-objects are located  
far enough from each other, they can exist independently and the system possess
the cluster property. This cluster property is 
important to the $N=2$ supersymmetry in the following sense. It is obvious
that the trace parts of
$A_{\mu}$ and $\psi$, or the parts proportional to identity matrix, are not
included in the action (\ref{S}), but play an essential role
in the $N=2$ supersymmetry transformations. This rather puzzling situation
may be resolved as follows. 
Due to the cluster property, the trace parts of diagonal-blocks become
collective coordinates and acquire the physical meaning 
as the centers of mass of 
the D-objects. In other words, space time coordinate is 
generated dynamically as the trace parts.
Therefore we are allowed to treat the trace parts of 
matrices.

\subsection{Determination of $\alpha$ and $\beta$}
\hspace*{\parindent}
In this subsection, we express the parameters $\alpha$ and $\beta$ in (\ref{S})
in terms of physical quantities using the results of the previous subsections.
Substituting one static D-string solution (\ref{Sstaticsolution}) into
(\ref{S}), we obtain
\beq
S=\alpha \frac{1}{4}(\frac{TL}{2\pi n})^2 n + \beta n
\label{nvariation}
\eeq
By taking the variation in terms $n$, we have
\beq
\frac{TL}{2\pi n}=2\sqrt{\frac{\beta}{\alpha}}
\label{relation1}
\eeq
as the equation of motion for $n$. 
Substituting this into (\ref{nvariation}), we obtain
\beq
S=\frac{1}{2\pi} \sqrt{\alpha\beta}TL
\eeq
This should be equal to the classical value of the Nambu-Goto action $\rho TL$, 
where $\rho$ is the string tension of the D-string.
Therefore we have
\beq
\sqrt{\alpha\beta}=2\pi \rho.
\label{relation2}
\eeq
We can also find the ratio of $\alpha$ and $\beta$ by examining the interaction
between two D-strings. As is well-known, $\rho$ and $\kappa$ are expressed
in terms of $\alpha^{'}$ and $g_s$ as
\beqa
\rho&=&4\pi^{5/2}\frac{\alpha^{'}}{\kappa} ,\n
\kappa&=&\gamma {\alpha^{'}}^2 g_s ,
\eeqa
where $2\kappa^2$ is $16$ times gravitation constant
and $g_s$ is the string coupling.
In terms of these quantities the effective action between
two anti-parallel static D-strings is expressed as
\beq
W=-2\rho^2 \frac{2\kappa^2}{4}(\frac{\pi^4}{3})^{-1}\frac{1}{b^6}TL,
\label{Wantiparalell2}
\eeq
where the factor $\pi^4/3$ comes from the area of $S^7$.
Requiring that this quantity is equal to 
(\ref{Wantiparalell}) and using (\ref{relation1}), we obtain a relation,
\beq
2\sqrt{3}\pi \alpha^{'}=\frac{TL}{2\pi n}=2\sqrt{\frac{\beta}{\alpha}}
\label{relation3}
\eeq
{}From (\ref{relation2}) and (\ref{relation3}), 
$\alpha$ and $\beta$ are determined as
\beqa
\alpha&=&\frac{8\pi^{\frac{5}{2}}}{\sqrt{3}\gamma}
                             \frac{1}{{\alpha^{'}}^2g_s}, \n
\beta &=&3\pi^2 \frac{8\pi^{\frac{5}{2}}}{\sqrt{3}\gamma} 
                             \frac{1}{g_s}.
\eeqa
Dropping an irrelevant numerical factor, we obtain
\beq
S=\frac{1}{{\alpha^{'}}^2 g_s}(- \frac{1}{4}Tr[A_{\mu},A_{\nu}]^2
            -\frac{1}{2}Tr(\bar{\psi}
           \Gamma^{\mu}[A_{\mu},\psi]))+3\pi^2 \frac{1}{g_s}
                                         Tr 1 .
\label{deteminedS}
\eeq

\vspace{1cm}

\section{The large-N reduced model of super Yang-Mills theory}
\setcounter{equation}{0}
\subsection{Double scaling limit}
\hspace*{\parindent}
In this section, we interpret the matrix model (\ref{S}) as an effective 
theory for the large-N reduced model of ten-dimensional super
Yang-Mills theory \cite{RM}. It is defined by
\beq
S_0=\frac{N a^4}{g_0^2}(-\frac{1}{4}Tr[A_{\mu},A_{\nu}]^2
                        -\frac{1}{2}Tr(\bar{\psi}\Gamma^{\mu}[A_{\mu},\psi])) ,
\label{S0}
\eeq
and
\beq
-\frac{\pi}{a} \leq \mbox{eigenvalues of }A_{\mu} < \frac{\pi}{a} ,
\label{eigenvaluerange}
\eeq
where $a$ is the space time cut-off.
In order to find a prescription of the double scaling limit, 
we consider ten-dimensional
super Yang-Mills theory before the large-N reduction:
 \beq
S_{SYM}=\frac{N}{g_0^2 a^6}\int d^{10}x (\frac{1}{4}Tr(F_{\mu\nu}F^{\mu\nu})
                                +\frac{i}{2}Tr(\bar{\psi}\Gamma^{\mu}
                                                            D_{\mu}\psi)),
\label{SSYM}
\eeq
where 
\beqa
F_{\mu\nu}&=&\partial_{\mu} A_{\nu}-\partial_{\nu} A_{\mu}+i[A_{\mu},A_{\nu}], \n
D_{\mu} \psi &=& \partial_{\mu} \psi + i[A_{\mu},\psi].
\eeqa
Since $a$ is the cut-off of length and $g_0$ is the dimensionless coupling 
constant, and
the theory possesses the classical mass scale 
$1/g_0^{\frac{1}{3}}a$. 
The quantization of this model poses some difficulty such as chiral anomaly.
However we use this model only to find the mass scale
of the large N reduced model (\ref{S0}) and (\ref{eigenvaluerange}).

Here we reinterpret the eigenvalues of $A_{\mu}$ as the space-time 
coordinates. Then $1/a$ becomes an infrared cut-off and the theory
possesses the mass scale $m=g_0^{\frac{1}{3}}a$.
As we will see in the next subsection, the path integrations of zero modes
give rise to
the term proportional $\beta$ in (\ref{S}), which we call 
the chemical potential term, and we can regard 
the matrix model (\ref{S}) as an effective theory of (\ref{S0}) and
(\ref{eigenvaluerange}). 
If we assume that we can naively set
\beq
\frac{N a^4}{g_0^2}=\alpha=\frac{4\pi^{\frac{5}{2}}}{\sqrt{6}\gamma}
                             \frac{1}{{\alpha^{'}}^2g_s} ,
\eeq
and
\beq
g_0^2 a^6=m^6,
\eeq
we obtain
\beq
Na^{10} \sim \frac{m^6}{{\alpha^{'}}^2 g_s} .
\eeq
We see then the double scaling limit should be taken such as
\beqa
a &\rightarrow& 0 ,\n
g_0 &\sim& a^{-3} \rightarrow \infty ,\n
N &\sim& a^{-10} \rightarrow \infty .
\eeqa

\subsection{Generation of the chemical potential term and 
                                 one-loop renormalization}
\hspace*{\parindent}
In this subsection, we consider the quantization of (\ref{S0}) and show that 
the matrix model (\ref{S}) can be regarded as its effective action. 
In particular we will see that the parameter $\beta$ in (\ref{S}) is
understood as a sort of chemical potential of the gas of eigenvalues of
$A_{\mu}$'s. We also examine the one-loop
renormalization and discuss that we should take the continuum limit of
the reduced model as the nonperturbative definition of the theory 
we are considering.

We first consider the path integral for $S_0$ (\ref{S0}) around a completely
diagonal background
\beq
p_{\mu}=\left(\begin{array}{cccc}
                d_{\mu}^{(1)}  &                &               &   \\      
                               & d_{\mu}^{(2)}  &               &   \\   
                               &                & \ddots        &   \\
                               &                &               &  d_{\mu}^{(N)}
                 \end{array} \right) .
\label{completediagonal}
\eeq
Here $d_{\mu}^{(i)}$'s are distributed uniformly from $-l$ to $l$, 
where $l$ is equal
to $\pi/a$ and corresponds to the infrared cut-off when we interpret $A_{\mu}$'s
as the space-time coordinates.
Then it is natural to interpret this background as the flat space-time.
We denote the partition function for this background as $Z_0(N)$.
As we have seen in section 3, the integrals of the off-diagonal elements cancel
each other between bosons and fermions. 
On the other hand, the diagonal elements do not appear in the one-loop level,
and $Z_0(N)$ is formally expressed as
\beqa
Z_0(N)&=&\int \prod_{i=1}^N\, \prod_{\mu=0}^9 \frac{dA_{\mu ii}}{S_N} \,
                              \prod_{\gamma=1}^{16}d\psi_{\gamma ii}\, 
                              dc_{ii}\, db_{ii}\; \n
      &=&\frac{1}{N!}l^{10N}\frac{1}{\mbox{vol(maximal torus)}}
         \int \prod_{i=1}^N\, \prod_{\gamma=1}^{16} d\psi_{\gamma ii}
\eeqa
The zero modes of the ghosts correspond to the maximal torus of the gauge group
whose action is trivial on the diagonal background (\ref{completediagonal}).
Therefore by dimensional analysis, it should be
\beq
\int \prod_{i=1}^N\,dc_{ii}\, db_{ii}\;=
\frac{1}{\mbox{vol(maximal torus)}}=const.\alpha^{\frac{N}{2}}.
\eeq
The zero mode integral of $\psi$ acquires non-zero value from
higher loop effects. 
Again by dimensional analysis, we should have
\beq
\int \prod_{i=1}^N\, \prod_{\gamma=1}^{16} d\psi_{\gamma ii}
\sim \alpha^{2N},
\eeq
and finally we obtain
\beq
Z_0(N)=const. \frac{1}{N!} (l^{10}\alpha^{\frac{5}{2}})^N.
\eeq
Next we consider the background 
\beq
p_{\mu}=\left(\begin{array}{cccc}
            \hat{A}_{\mu}  &                &               &   \\      
                           & d_{\mu}^{(1)}  &               &   \\   
                           &                & \ddots        &   \\
                           &                &               &  d_{\mu}^{(N-n)}
                 \end{array} \right) 
\eeq
Here $\hat{A}_{\mu}$ is a $n \times n$ matrix and $d_{\mu}^{i}$'s are distributed
uniformly from $-l$ to $l$. Denoting the partition function around this
background as $Z$, we define the effective action for the $n \times n$ block as
\beq
S_{eff}=-\log \frac{Z}{Z_0(N)}.
\eeq
Noting that the contribution from the lower right $(N-n) \times (N-n)$ 
block is nothing but
$Z_0(N-n)$ and the contributions from the off-diagonal-blocks are calculated
in section 3.4. We obtain
\beqa
S_{eff}&=&-\log\frac{Z_0(N-n)}{Z_0(N)}  
          -\alpha \frac{1}{4}Tr[\hat{A}_{\mu},\hat{A}_{\nu}]^2                             + const.\sum_{i=1}^{N-n}\frac{1}{(d^{(i)})^8}
                          Tr(\hat{F}\hat{F}\hat{F}\hat{F})
       +\cdots \n
       &=& n \log(\frac{l^{10}\alpha^{\frac{5}{2}}}{N})
                -\alpha \frac{1}{4}Tr[\hat{A}_{\mu},\hat{A}_{\nu}]^2
           + const.\frac{N}{l^{10}}l^2 Tr(\hat{F}\hat{F}\hat{F}\hat{F})
          +\cdots ,\n
\eeqa
where $\hat{F} \sim i[\hat{A}_{\mu},\hat{A}_{\nu}]$. As was shown 
in the previous subsection, in the double scaling limit, 
$N \sim l^{10}$. Therefore we have
\beq
S_{eff}= const. n -\alpha \frac{1}{4}Tr[\hat{A}_{\mu},\hat{A}_{\nu}]^2
         + const. \frac{1}{a^2}Tr(\hat{F}\hat{F}\hat{F}\hat{F})
          +\cdots .
\eeq
The first term is the chemical potential term we are looking for.
The third term and the following terms are cut-off dependent and divergent. 
This means that we have to add counter terms to $S_0$ (\ref{S0}).
As is well-known in the large-N gauge theory, the structures 
of these divergences are the same as those of the original
ten-dimensional super Yang-Mills theory (\ref{SSYM}) \cite{RM}, which is not 
renormalizable perturbatively. We cannot give a complete definition of
the theory (\ref{S0}) only by the perturbation theory
as we cannot do so for the super Yang-Mills theory (\ref{SSYM}). 
Namely we expect
that there is a nontrivial fixed point of renormalization group of the theory
(\ref{S0}) and we can take the continuum limit around it to define the theory 
nonperturbatively. The divergences of the loop corrections are 
mild due to supersymmetry, which supports the existence of the nontrivial 
fixed point. At the one loop level the degree of the divergences is
less than or equal to two.
There would be the sixth order divergences without supersymmetry.
We also expect that the matrix model we consider is
controlled by a universality similar to field theory.

\subsection{Moduli space of the matrix model}
\hspace*{\parindent}
Our matrix model possesses a large number of degenerate vacua at least
perturbatively. In this subsection, we consider the following classical 
solution:
\beqa
\bar{A}_0 &=&\left(\begin{array}{cccc}
          d_0^{(1)}  &                &               &   \\      
                               & d_0^{(2)}  &               &   \\   
                               &                & \ddots        &   \\
                               &                &               &  d_0^{(N)}
                 \end{array} \right) ,\n
\bar{A}_i&=&0, 
\eeqa
where $d_0^{(i)}$'s are distributed uniformly from $0$ to $T$.
At this point of the moduli space, our action (\ref{S0}) becomes the following
in the $A_0=\bar{A_0}$ gauge:
\beq
\int_0^T d\tau \frac{1}{2}[-Tr(\frac{\partial A_i}{\partial \tau})^2
-\frac{1}{2}Tr[A_i,A_j]^2 
-iTr\psi\partial_{\tau}\psi
-Tr\bar{\psi}\Gamma^i[A_i,\psi]] .
\eeq
Here we have put $[\bar{A}_0, A_i]=i\partial_\tau A_i,~
[\bar{A}_0,\psi]=i\partial_\tau \psi$
which can be justified in the large-N limit \cite{RM}.
After the Wick rotation $(\tau\rightarrow i t)$, this action can be seen to 
coincide with that of Banks, Fischler, Shenker and Susskind \cite{BFSS}:
\beq
\frac{1}{2g}[Tr\dot{X}^i\dot{X}^i+2Tr\bar{\theta}\gamma_{-}\dot{\theta}
-\frac{1}{2}Tr[X^i,X^j]^2-2Tr\bar{\theta}\gamma_{-}\gamma^i[\theta,X^i]]
\eeq
The fermionic variables are $32$ component eleven-dimensional spinors, 
satisfying the light-cone constraint $\gamma_{+}\theta=0$.
They have proposed this action as a nonperturbative formulation of M-theory
in the light-cone frame. Therefore we found that our matrix model contains
type IIA string theory as well! 

However we have always regard our matrix model as a nonperturbative definition
of type IIB string theory in this paper. In section 3, we have identified 
particular moduli which can be interpreted in terms of type IIB string theory.
However these findings are not mutually contradictory since type IIA and type IIB
string theories are T-dual to each other.

Therefore we can view the finding of this section as another manifestation of 
duality. Namely the different points of the moduli space of our matrix model 
may be interpreted either by type IIB string theory or by type IIA string theory.
This feature is also very reminiscent of $N=2$ supersymmetric gauge theories.

\vspace{1cm}

\section{Summary and discussion}
\setcounter{equation}{0}
\hspace*{\parindent}
In this paper, we proposed a matrix model which may give a constructive 
definition of string theory. We showed that the matrix model is connected
to the type IIB Green-Schwarz action and possesses the $N=2$ supersymmetry,
which is a sufficient condition for the theory to include gravitation, 
and that the space-time
is dynamically generated via the collective coordinate of this model.
We also showed that
the massless spectrum is consistent with string theory by examining the
interactions between the BPS-saturated states. We conjectured that
the theory can be defined nonperturbatively
by the continuum limit of the large-N reduced model of ten-dimensional
super Yang-Mills theory, whose existence is expected due to the supersymmetry,
and pointed out that the marginal preservation of the $U(1)^d$ symmetry is 
important for the generation of space time.

Finally we discuss the relation between the Wilson loop and the classical
solutions for $A_{\mu}$'s. 
The Wilson loop of the large-N reduced model is given by
\beq
W[\xi(\sigma)]=TrP\exp (i \oint_0^{2\pi} 
                           A_{\mu} \dot{\xi}^{\mu}(\sigma) d\sigma),  
\label{Wilsonloop}
\eeq
where $\xi^{\mu}(\sigma)$ satisfies
\beq
\xi^{\mu}(2\pi)=\xi^{\mu}(0)+n^{\mu}a,
\eeq
and the $n^{\mu}$'s are integers and correspond to the winding numbers.
As in the previous sections if
we regard $A_{\mu}$'s as the space-time coordinates,
$\dot{\xi}^{\mu}$  corresponds to the momentum. 
Therefore it is natural to consider
the Wilson loop (\ref{Wilsonloop}) as the creation operator of the string state
with momentum
eigenvalues $|k^{\mu}(\sigma)=\dot{\xi}^{\mu}(\sigma)\rangle$.
If we identify the Wilson loops with strings, then it is natural to identify
the classical solutions for $A_{\mu}$'s with D-objects. In order to see that
it is really the case, 
we again consider the static D-string solution:
\beqa
p^D_{0}&=&\left(\begin{array}{cccc}
           \frac{T}{\sqrt{2\pi n}}q &                &               &   \\      
                           & d_{0}^{(1)}  &               &   \\   
                           &                & \ddots        &   \\
                           &                &               &  d_{0}^{(N-n)}
                 \end{array} \right) ,\n 
p^D_{1}&=&\left(\begin{array}{cccc}
           \frac{L}{\sqrt{2\pi n}}p &                &               &   \\      
                           & d_{1}^{(1)}  &               &   \\   
                           &                & \ddots        &   \\
                           &                &               &  d_{1}^{(N-n)}
                 \end{array} \right) ,\n
p^D_{i}&=&\left(\begin{array}{cccc}
                   a_i 1_n &                &               &   \\      
                           & d_{i}^{(1)}  &               &   \\   
                           &                & \ddots        &   \\
                           &                &               &  d_{i}^{(N-n)}
                 \end{array} \right) .
\label{Dstringmatrix}
\eeqa
Here the first block represents a D-string parallel to the $X^0-X^1$ plane and
located at $X^i=a^i$, and the rest represents the flat vacuum. Therefore
$p^D_{\mu}$ represents one D-string sitting in the vacuum, while the flat vacuum
is expressed as
\beq
p^{vac}_{\mu}=\left(\begin{array}{cccc}
                d_{\mu}^{(1)}  &                &               &   \\      
                               & d_{\mu}^{(2)}  &               &   \\   
                               &                & \ddots        &   \\
                               &                &               &  d_{\mu}^{(N)}
                 \end{array} \right) .
\label{vacuummatrix}
\eeq
The difference of the Wilson loop for these backgrounds is given 
in the tree level by
\beq
<W(k^{\mu}(\sigma))>_{A_{\mu}=p^D_{\mu}}
-<W(k^{\mu}(\sigma))>_{A_{\mu}=p^{vac}_{\mu}}
=e^{ik_0^i a^i}TrP\exp[i\oint (\frac{T}{2\pi n}q k^0(\sigma)
                              +\frac{L}{2\pi n}p k^1(\sigma))d\sigma].
\label{Wilsonloopvev}
\eeq
If the Wilson loops correspond to strings and the classical solutions
correspond to D-strings, this quantity should be compared with the closed string
tadpole wave function in the presence of the D-string.
Denoting the boundary state corresponding to the D-string as $|D\rangle$,
the closed tadpole wave function is given by
\beq
\langle k(\sigma)|D\rangle 
= \delta(k^0(\sigma))\delta(k^1(\sigma)) e^{ik_0^i a^i} .
\label{Dstringwavefunction}
\eeq
The exponential factor representing the Dirichlet boundary condition in 
(\ref{Dstringwavefunction}) is reproduced in (\ref{Wilsonloopvev}).
Since the Wilson loop does not create an on-shell state, we cannot demand
the exact coincidence. At any rate the interpretation of the Wilson loops and
D-strings seems consistent.
This is the reason why we have called the classical solutions as D-strings.

\vspace{1cm} 

We would like to thank H. Aoki, K.-J. Hamada, T. Tada and
K. Yoshida for discussions and J.H. Schwarz for valuable comments.

\vspace{0.5cm}

\noindent
\begin{large}
Note added
\end{large}

After finishing this paper, we have been informed that Jevicki and Yoneya 
\cite{JY}, and Periwal \cite{Periwal} have considered related models.

\newpage

\section*{Appendix}
\renewcommand{\theequation}{A.\arabic{equation}}
\hspace*{\parindent}
In this appendix, we calculate the one-loop quantum corrections 
of the matrix models (\ref{quantumS}) and (\ref{S0}). As in the ordinary
background field method in the quantum field theories, we decompose the matrices
$A_{\mu}$ and $\psi$ to the backgrounds and the quantum fluctuations. Namely,
\beqa
A_{\mu} &=& p_{\mu}+a_{\mu},\\
\psi &=& \chi + \varphi ,
\eeqa
where $p_{\mu}$ and $\chi$ are backgrounds and $a_{\mu}$ and $\varphi$ are
quantum fluctuations. The action is expanded up to the second order of the
quantum fluctuations such as
\beqa
 S_2 &=& -Tr ( \frac{1}{4}[p_{\mu},p_{\nu}]^2
            +\frac{1}{2}\bar{\chi}
           \Gamma^{\mu}[p_{\mu},\chi] \n
    & & -a_{\nu}([p_{\mu},[p_{\mu},p_{\nu}]]
          +\frac{1}{2}[\bar{\chi}\Gamma^{\mu},
           \chi])+\bar{\psi}\Gamma^{\mu}[p_{\mu},\varphi] \n
    & &+ \frac{1}{2}{[p_{\mu},a_{\nu}]}^2-\frac{1}{2}{[p_{\mu},a_{\mu}]}^2
          +[p_{\mu},p_{\nu}][a_{\mu},a_{\nu}]  \n
    & &+ \frac{1}{2}\bar{\varphi}\Gamma^{\mu}[p_{\mu},\varphi]
      +\bar{\chi}\Gamma^{\mu}[a_{\mu},\varphi] ) ,
\label{S2}
\eeqa
where the terms in the second line vanishes due to the equations of motions, or
are dropped in the back ground field method.
To fix the gauge invariance
\beqa
\delta A_{\mu}&=&i[A_{\mu},\alpha],\n
\delta\psi&=&i[\psi,\alpha],
\eeqa
we add the gauge fixing term 
\beq
S_{g.f.}=-Tr(\frac{1}{2}[p_{\mu},a_{\mu}]^2+[p_{\mu},b][p_{\mu},c]),
\label{gaugefixingterm}
\eeq
where $c$ and $b$ are ghosts and anti-ghosts, respectively. In the following,
we set $\chi=0$. Dropping the first order term of the quantum fluctuations,
we obtain 
\beqa
\tilde{S}_2=S_2+S_{g.f}=-Tr&(& \frac{1}{2}[p_{\mu},a_{\nu}]^2
                          +[p_{\mu},p_{\nu}][a_{\mu},a_{\nu}] \n
                    &+& \frac{1}{2} \bar{\varphi}\Gamma^{\mu}[p_{\mu},\varphi]
                   +  [p_{\mu},b][p_{\mu},c]).
\label{S2tilde}
\eeqa
Here we introduce a notation of adjoint operators. For an operator $o$,
\beq
[o,X]=OX.
\eeq
Using this, we rewrite (\ref{S2tilde}) as
\beq
\tilde{S}_2=Tr(\frac{1}{2}a_{\mu}(P_{\lambda}^2 \delta_{\mu\nu}
                      -2iF_{\mu\nu})a_{\nu}-\frac{1}{2}\bar{\varphi}\Gamma^{\mu}
                      P_{\mu}\varphi+b P_{\lambda}^2 c) ,
\eeq
where
\begin{eqnarray}
        [p_{\mu},X] & = & P_{\mu}X,\\
       \left[ f_{\mu \nu},X \right] & = & F_{\mu \nu}X,
          \mbox{} f_{\mu \nu}=i[p_{\mu},p_{\nu}] .
\end{eqnarray}
{}From this, the one-loop effective action $W$ is evaluated as
\beqa
W &=& -\log \int da d\varphi dc db \,e^{-\tilde{S}_2} \n
  &=& \frac{1}{2}Tr \log(P_{\lambda}^2 \delta_{\mu\nu}-2iF_{\mu\nu})
      -\frac{1}{4}Tr \log((P_{\lambda}^2+\frac{i}{2}F_{\mu\nu}\Gamma^{\mu\nu})
      (\frac{1+\Gamma_{11}}{2})) \n
  & & -Tr \log(P_{\lambda}^2)+i\theta ,
\eeqa
where $\theta$ is the anomaly term. We can show that $\theta$ vanishes for
the two cases we have studied in this paper.
\beq
\mbox{Case(1):}\: F_{\mu\nu}=0.
\eeq
This is because we can simultaneously diagonalize $P_{\mu}$'s and 
$\Gamma^{\mu} P_{\mu}$ becomes a real operator.
\beq
\mbox{Case(2):} \: P_i=0 \:\:\mbox{for at least in one direction}\: i.
\eeq
This is because 
\beq
\det( \Gamma^{\mu} P_{\mu})^{\dagger}(\Gamma^{\nu}P_{\nu})
=\det(\Gamma^i \Gamma^{\mu} P_{\mu})^2.
\eeq

%
%
%

\end{document}